\documentclass[11pt,a4]{article}
\usepackage[margin=1.0in]{geometry}
\usepackage{graphicx}
\usepackage{epsfig}
\usepackage{amssymb,amsfonts}
\usepackage{mathrsfs}
\usepackage{epstopdf}
\usepackage{hyperref,color}
\usepackage{natbib}
\usepackage{setspace}
\usepackage{xypic}\usepackage[all,cmtip]{xy}

\usepackage{braket}
\usepackage{amsmath}

\begin{document}

\title{Laws beyond spacetime}
\author{Vincent Lam\thanks{Institute of Philosophy, University of Bern, CH-3012 Bern, Switzerland. Email: vincent.lam@unibe.ch}
\thanks{School of Historical and Philosophical Inquiry, The University of Queensland, St Lucia QLD 4072, Australia. Email: v.lam@uq.edu.au} 
and Christian W\"uthrich\thanks{Department of Philosophy, University of Geneva, CH-1211 Geneva, Switzerland. \mbox{Email: christian.wuthrich@unige.ch}}
\thanks{We thank Nick Huggett and Carl Hoefer for fruitful discussions, and audiences at {\em Foundations 2018} in Utrecht, Geneva (live and over zoom), and Pittsburgh (over zoom) for their engagement with the project. Work on this project has been supported in part by the Swiss National Science Foundation through project 105212$\_$169313 and by the John Templeton Foundation through the project \textit{Cosmology Beyond Spacetime}.}
}

\date{}

\maketitle

\begin{abstract}
\noindent
Quantum gravity's suggestion that spacetime may be emergent and so only exist contingently would force a radical reconception of extant analyses of laws of nature. Humeanism presupposes a spatiotemporal mosaic of particular matters of fact on which laws supervene; primitivism and dispositionalism conceive of the action of primitive laws or of dispositions as a process of `nomic production' unfolding over time. We show how the Humean supervenience basis of non-modal facts and primitivist or dispositionalist accounts of nomic production can be reconceived, avoiding a reliance on fundamental spacetime. However, it is unclear that naturalistic forms of Humeanism can maintain their commitment to there being no necessary connections among distinct entities. Furthermore, non-temporal conceptions of production render this central concept more elusive than before. In fact, the challenges run so deep that the survival of the investigated analyses into the era of quantum gravity is questionable.
\end{abstract}



\section{Introduction}
\label{sec:intro}

What are laws of nature? Adopting a naturalistic attitude, this question assumes a new urgency in light of tentative results in fundamental physics, or so we will argue. A naturalistic attitude involves a close consideration of the adequacy of a proposed analysis of laws in light of scientific results and practices. The naturalism we will adopt at least requires that our metaphysical theories be consistent with results in the empirical sciences. 

A truly fundamental theory of physics will include a quantum theory of gravity. Current fundamental physics rests on the standard model of particle physics and general relativity. Neither of them can be a truly fundamental theory, as neither is universal: black holes and the very early universe remain outside the remit of general relativity, and the standard model does not accommodate gravity and so the dynamical geometry of spacetime. A theory of quantum gravity is expected to remedy the former, and perhaps to address the latter. To date, there exists an impressive diversity of research programs aiming at delivering such a theory. They differ significantly in their basic principles, methods, and results. But they all have one thing in common: none of them enjoys empirical support beyond the support of the `old' physics they may incorporate. 

Perhaps surprising given their divergencies, one finds a near consensus among the different approaches to quantum gravity according to which the fundamental structure is significantly non-spatiotemporal. This denial of the fundamentality of spacetime comes in different degrees and manifests itself in different ways in different research programs.\footnote{\citet{hugwut13}; for a detailed account of how spacetime disappears and re-emerges in major approaches to quantum gravity, see \citeauthor{OON} (forthcoming). In causal set theory, any spatial structure whatsoever is emergent \citep{Wuthrich2020}; in group field theory, there is no spatial extension fundamentally, see \citeauthor{Lam-Oritiforthcoming} (forthcoming); loop quantum gravity faces the `problem of time' according to which time, or at least change, must emerge \citep{hugeal13}; and it has been argued, convincingly to our mind, that spacetime is also emergent in string theory \citep{Huggett2015}.} On these approaches, spacetime is not fundamental, but merely emergent. The disappearance and re-emergence of spacetime thus appears to be a generic consequence of quantum gravity. 

If this is right, then an analysis of laws of nature cannot depend on spacetime.\footnote{At least an analysis of \textit{fundamental} laws cannot depend on spacetime---one of derivative laws may be untouched by quantum gravity.} As extant analyses of laws of nature appear to presuppose the (necessary) existence of spacetime, they thus stand in tension with this suggestion from quantum gravity. This paper offers an analysis of the deep ramifications of the emergence of spacetime for three major analyses of laws: the Humean best systems analysis (in \S\ref{sec:hume}), primitivism, and dispositionalism (both in \S\ref{sec:pd}). 

Before we delve into the discussion of the ramifications for particular analyses of laws, let us address one general worry: one might object to our project by insisting that the mere \emph{existence} of spacetime---as opposed to its fundamentality---suffices for the analyses of laws to accommodate whatever dependence they may have on spacetime. However, this would not entirely take care of the concerns that we are about to raise. A theory of quantum gravity need not take the emergence of spacetime to be necessary, i.e., as emergent in all circumstances judged physically possible by the theory; instead, it suffices to establish that it permits the emergence of spacetime, possibly only under fortuitous conditions. Thus, an analysis depending on the existence of spacetime would fail at those worlds where circumstances would be insufficiently auspicious for spacetime to emerge. A naturalist analysis of laws should work under all conditions deemed possible by our best scientific theories. This means, in particular, that it needs to cover all possibilities licenced by fundamental physics, including those in which spacetime may be absent. Moreover, to the extent that research programs in quantum gravity describe fundamental non-spatiotemporal structures in terms of certain (novel) physical laws, any serious account of fundamental physical laws of nature should apply to these latter as well---even in cases where spacetime does not emerge from these fundamental non-spatiotemporal structures.

Our thesis is that the emergence, i.e., the non-fundamentality, of spacetime would have serious implications for our conceptions of what it is to be a law of nature, implications that we try to articulate in this essay. As stated, \S\ref{sec:hume} discusses the best systems approach to laws and Humeanism more broadly. \S\ref{sec:pd} deals with primitivism about laws and with dispositionalism. Conclusions follow in \S\ref{sec:conc}. Please note that for the purposes of this paper, we accept the possibility of a fundamentally non-spatiotemporal world although the science remains of course tentative. Everything we say thus remains under the presupposition that the non-fundamentality of spacetime is borne out by our best science. 

\section{Best systems without spacetime}
\label{sec:hume}

How would a Humean account of laws of nature look like if spacetime is at best emergent, rather than fundamental? Would such an approach to analysing laws even be viable in such a world? In this section, we will first (in \S\ref{ssec:hneg}) try to sketch the negative consequences for what we take to be the standard Humean approach. As we hope this part will establish, the potential absence of spacetime requires radical revisions of the Humean approach, at least as compared to extant versions. In \S\ref{ssec:hpos}, we outline what we take to be essential elements for a Humean analysis of laws revised and generalized in view of the possibility of merely emergent spacetime.

\subsection{Humean supervenience and spacetime}
\label{ssec:hneg}

Humean approaches to laws of nature earn their epithet by denying that there exist `necessary connections' between distinct entities. By far the best worked out and most popular Humean analysis of laws of nature is the so-called `best systems analysis' developed in \citet{Ramsey1978}, \citet[\S3.3]{Lewis1973}, \citet{Lewis1983}, and \citet{Earman1984}.\footnote{For a more recent take on the Humean reductionist programme, see \citet{Hall2015}. What we describe below is what Hall calls the `official' Humean reductionist story, as opposed to the `unofficial' one according to which the best system is determined by the ``epistemic standards \textit{for determining the nomological possibilities for the non-modal facts}'' (266) which are operative in science.} 

According to this analysis, a proposition is a law of nature just in case it is an axiom or a theorem of the `best system'. The best system is a true, deductively closed theory which best balances simplicity in expression with strength in explanation. How to `best' `balance' `simplicity' and `strength' is the subject of long-standing debates about the virtues and vices of the best systems analysis. As we wish to articulate a problem for the account beyond the entrenched lines, we are happy to grant, for present purposes, that there exist objectively best systems of the requisite kind. With this generous allowance, the analysis gets at least off the ground.

Or so it might seem at first. But let us have a closer look at what happens in case spacetime turns out to be merely emergent. Prima facie, however, the non-fundamentality of spacetime poses no threat to the best systems analysis, which is characterized, here and elsewhere, without direct reference to spacetime and so might appear untouched by a possible absence of spacetime. In other words, the advocate of the best systems analysis might simply formulate whatever theory is best according to the relevant criteria, read off what the laws of nature are---spatiotemporal or not---, and get on with business. In this general procedure, no reference is made to spacetime and no proscription is imposed that the laws be dynamical (and so spatiotemporal). Thus, the analysis seems sufficiently general to accommodate the possibility that the best theory does not postulate or imply the (fundamental) existence of spacetime.

But matters are not quite as straightforward as that. The larger Humean project is to reduce modal notions such as possibility and necessity, and consequently of lawhood, to fundamental occurrent, modally and nomically innocent, facts.\footnote{Although our argument does not depend on a specific conception of `facts', we conceive of facts as `states of affairs', i.e., an object exemplifying a property or one or more objects standing in a relation. `Occurrent facts', on our use, are states of affairs which obtain, or are the case.} After all, there are not supposed to be any necessary connections in the world. Thus, necessity, including nomic necessity, can only be tolerated if it is derivative and non-fundamental. Therefore, the Humean project consists in two distinct steps: first, one identifies a non-modal basis, which serves to express the occurrent facts to which modality gets reduced; and second, one adds a recipe to reduce modal notions such as laws of nature to this basis. The best systems analysis concerns the second step. But even so, it requires that the first step also be completed in order for the recipe to find its application. Consequently, even if the \textit{recipe} to derive the laws from the fundamentalia does not itself depend in any way on there being spacetime, the approach still faces the novel challenge if the basis---the fundamentalia---rely on, and so include, spacetime. 

Let us look at an example in order to appreciate the point. In the Humean camp, the best systems analysis offered by Lewis is arguably the most systematic and ambitious project in that family. Lewis's reduction takes \textit{worlds} as fundamental. A metaphysically possible world, for Lewis, consists in a spatiotemporal arrangement of local qualities, i.e., ``perfectly natural properties which need nothing bigger than a point at which to be instantiated'' \citep[x]{Lewis1986}. These points may be parts of spacetime itself or ``point-sized bits of matter or aether or fields'' (ibid.), or both. Even in the latter case, however, there is ``a system of external relations of spatiotemporal distance between points'' (ibid.). We may be substantivalists or relationalists about it, but one way or the other, spacetime forms the backbone of a Lewisian world. For Lewis, spatiotemporal relations are what unify and isolate the worlds so central to his analysis. They are the `world-making' relations, i.e., the relations in which all and only worldmates stand. Spacetime thus furnishes the backbone of Lewis's project.\footnote{\label{fn:analogical}At this point, an important caveat ought to be stated: Lewis clearly acknowledges the potential need to generalise the world-making relations from spatiotemporal relations to what he dubs `analogically spatiotemporal' relations \citep[75f]{LewisPlurality}. These relations are characterized by four conditions. First, they are \textit{natural}. Second, they are \textit{pervasive}, i.e., if there is a chain of relations connecting a `first' to a `last' object, then there is also a direct relation between them. Third, they are \textit{discriminating}, i.e., the relational structures uniquely identify the positions in them which the relata can occupy, such that the relata are identifiable by their positions only. Fourth, they are \textit{external}, i.e., ``they do not supervene on the intrinsic natures of the relata taken separately, but only on the intrinsic character of the composite of the relata'' \citep[76]{LewisPlurality}. The condition of relations qualifying as analogically spatiotemporal is strictly weaker than as spatiotemporal relations, as the latter are also analogically spatiotemporal but not vice versa. For an argument why we should not in general expect the fundamental relations in quantum gravity to be even analogically spatiotemporal, see \citet{Wuthrich2020}. \citet{Darby2009} argues that the non-local relations sometimes obtaining between quantum-mechanical systems are neither pervasive nor discriminating and so not even analogically spatiotemporal. He thus pleads to further weaken the requirement and to take two objects as worldmates if and only if they stand in a natural external relation (see also \citealt{Bricker1996} for an earlier argument that Lewis's criterion should be relaxed to a mere demand of the externality of the worldmate relation). Arguably, the (non-local) relations quantum-mechanical systems can stand in are natural and external, and can thus function as world-making relations. We concur with Darby that the demand for analogically spatiotemporal relations ought to be weakened to the wider category of natural external relations, also in light of our analysis. But we are getting ahead of ourselves.}

``All else'', Lewis tells us (ibid.), including, importantly, laws of nature, supervenes on the arrangement of local qualities at a world. This is Lewis's doctrine of \textit{Humean supervenience}. \citet[51]{Maudlin2007} usefully identifies two logically independent doctrines which jointly amount to Humean supervenience:
\begin{description}
\item[Separability:] ``The complete physical state of the world is determined by (supervenes on) the intrinsic state of each spacetime point (or each pointlike object) and the spatio-temporal relations between those points.'' (ibid.)
\item[Physical Statism:] ``All facts about the world, including modal and nomological facts, are determined by its total physical state.'' (ibid.)
\end{description}
We take Physical Statism to be a version of what it \textit{is} to be Humean. It can be found, in one form or another also in Mill, Ramsey, and Earman. It captures the idea that necessity and lawhood are derivative of, and secondary to, the totality of occurrent facts. Separability, on the other hand, articulates the idea that the world is local in the sense that the state of affairs at any particular location in spacetime does not depend on the state of affairs elsewhere in the spacetime continuum. 

Humeanism starts out from the actual world and its basis of fundamental, non-modal facts. From this vantage point, a principle of `recombination' is used to span a range of (metaphysical) possibility. According to this principle, any distribution of the perfectly natural properties in spacetime (and perhaps all different ways spacetime itself can be arranged) is possible. In other words, the local, intrinsic states of affairs are freely recombinable into novel possibilities, spanning the vast Lewisian pluriverse of all metaphysically possible worlds. At each of these worlds, the laws of nature are determined by the best system at that world, such that worlds consistent with those laws are the `nomically possible' worlds.\footnote{Another important caveat: \citet[x]{Lewis1986} himself takes Humean supervenience to be a contingent truth, as there might be ``unHumean'' worlds containing `alien' properties such that Humean supervenience fails. The best systems analysis may fail at those worlds.} Thus, although the general template for a best systems analysis does not depend on spacetime, the Lewisian version certainly does in its construction of possible worlds and hence of the pluriverse, which forms the basis of Lewis's system.\footnote{This is a necessary truth, as even the unHumean worlds are unified and isolated by spatiotemporal relations.} In fact, for any reductive project such as Lewis's, a central task is to characterize the supervenience basis on which the laws are supposed to supervene. In light of the challenge articulated in \S\ref{sec:intro}, the central question becomes whether or not this supervenience basis contains spacetime or spatiotemporal relations. For Lewis, it clearly does.

Lewis's Humean system has been criticised from many directions. For instance, \citet[ch.\ 2]{Maudlin2007} offers an influential critique of Humean supervenience, rejecting Humean supervenience because he takes Separability to be contradicted by physical theory and Physical Statism to run contrary to scientific practice. The central naturalistic objection to Humean supervenience is that it appears to violate quantum non-locality as found in the correlations of entangled bipartite quantum systems. The total state of these systems does in general not supervene on the states of their localisable parts. In the literature, this has been interpreted as a sign of a form of holism in quantum physics \citep{Teller1986,Healey1991}. In fact, Lewis himself was well aware of the problem, which, however, he dismissed by famously insisting that he is not prepared ``to take lessons in ontology from quantum physics'' before ``it is purified of instrumentalist frivolity... of doublethinking deviant logic... and... of supernatural tales about the power of the observant mind to make things jump'' \citep[xi]{Lewis1986}. Although we concur with Lewis that these are serious points in need of being addressed by any viable interpretation of quantum physics, quantum non-locality remains present in realist and physicalist interpretations properly ``purified'' as demanded by Lewis. Thus, quantum non-locality does put serious pressure on Humean supervenience.\footnote{Lewis was much less dismissive of quantum mechanics in later work, cf.\ \citet{Lewis1994,Lewis2004}. In the latter paper, Lewis expresses confidence that the defence of Humean supervenience ``can doubtless be adapted to whatever better supervenience thesis may emerge from better physics'' (474). This article challenges his confidence.}

There is no need to pursue all the wrinkles of the ensuing debate regarding the tenability of Humean supervenience in light of quantum non-locality. However, for our present concern, it will be illuminating to briefly mention two Humean responses to the problem.\footnote{We exclude from present consideration any approach which includes superluminal signalling, backward causation, superdeterminism, or other radical revisions of standard physics and the standard way of thinking about it. In other words, we accept the existence of quantum non-locality.} Both of these responses involve more or less severe modifications of Lewis's theory, and so may shed light onto the directions a Humean response to the problem from quantum gravity could take. A straightforward response, proposed e.g.\ in \citet{Darby2012}, to the challenge is to include the non-local physical relations between point-sized entities into the supervenience base. Darby understands that a move to include such relations will render the supervenience base less `local' and more `global': instead of being restricted to particular matters of fact localisable to points, the base now includes global matters of particular fact.\footnote{`Global' facts are opposed to `local' ones, while `particular' facts stand in contrast with `general' ones. Arguably, these distinctions are orthogonal to one another.} Locality is lost, but supervenience is saved. This defence opts to replace Lewis's version of Separability by relaxing to a more flexible supervenience base of the total physical state of the world. The locality present in the original proposal by Lewis was supposed to offer, we take it, a natural ground on which to exercise the generation of alternative possibilities through free recombination. Locality as such is not essential to the Humean project and so can be given up; free recombination, on the other hand, is what is supposed to guarantee the absence of necessary connections and so is absolutely vital to it. Let us heed the lesson emerging here: however alien to the original proposal the supervenience base may become in light of new physics, it must remain amenable to free recombination or some equivalent means of ascertaining the absence of necessary connections. 

The second response, due to \citet[104]{Loewer1996}, saves a form of locality (and so of separability) by replacing fundamental physical space with configuration space.\footnote{Loewer offers his solution in the context of Bohmian mechanics, even though the general strategy applies more widely.} On this view, the ontology consists in point particles, or rather in their spatial degrees of freedom, and in time. This naturally leads to a space of $3n$ dimensions for $n$ fundamental particles which can move in three spatial dimensions. The wave functions of quantum mechanics are functions defined on such configuration spaces (plus time). The total state of all $n$ particles can thus be captured by a single point moving in configuration space over time. Thus, the supervenience basis consists not in intrinsic states of points in spacetime, but instead of point-like matters of particular fact in configuration space. The basis will need to include the relations among the points in configuration space.\footnote{\citet{Maudlin2007} complains that this move elevates Separability to the level of a regulative principle, rather than a result of empirical enquiry. As naturalists, we concur that metaphysical principles, including regulative ones, cannot ultimately be beyond the reach of empirical disconfirmation, however indirect. But the purpose of our present project is to determine the extent to which Humeanism is under pressure from quantum gravity, so we will not press the objection from quantum non-locality.} 

It remains an interesting interpretative question whether spacetime vanishes, or more generally, what the metaphysical connection between configuration space and physical space(time) is on this approach. It is not obvious that space is entirely absent, as the metaphysical vantage point of the approach was to posit $n$ particles in three-dimensional physical space. It is unclear what we are to make of that stipulation if spacetime does not exist. Should we perhaps treat the stipulation as a heuristic without ontic implications? And time is clearly still part of the ontology. Be this as it may, in \S\ref{ssec:hpos}, we will see that the approach might nevertheless offer a clue as to how to deal with the disappearance of spacetime. 

In sum, although the locality of basal matters of facts, and hence Separability, was attractive to the Humean due to its natural connection to free recombination, we see no in principle obstacle of generalizing the basic mosaic of occurrent facts to include non-local or relational facts in the supervenience basis. However, Humean strategies must more radically adapt once spacetime is recognized to be emergent rather than fundamental. An altogether different way of characterizing the basis is required in case spacetime is not fundamental. As argued in \S\ref{sec:intro}, this means that any analysis of laws which seeks to be applicable to all possible ways in which the world could be cannot rely on the existence of spacetime. For the Humean, this means that Humean supervenience, and in particular Separability, must either be reformulated in ways to make it independent of spacetime, or else be abandoned altogether. 

In fact, if spacetime is emergent, then not only must Separability be reformulated, but Lewis's popular figure of speech of the Humean mosaic becomes misleading, and the objection from quantum non-locality cannot even be properly stated anymore. The very metaphor of the mosaic invokes an image as of local occurrent \emph{and spatially arranged} matters of particular fact. If the spatial arrangement of the occurrent facts forming the supervenience base merely emerges, and only contingently so, a new metaphor must be devised in order to describe the new base. We are at a loss concerning an adequate metaphor, but prefer the image of a carpet or a fabric where the non-point-like parts are interwoven with connecting threads to form a whole. We admit that both a carpet or a fabric are still highly suggestive of a spatial organization, but at least their spatial arrangements are pliable, flexible, and dynamical, and not rigid and fixed as in the case of a mosaic.  In light of the dynamic spacetimes of general relativity, the carpet is certainly a superior metaphor to the mosaic. Moving to the context of QG, we insist that these connecting threads be understood as non-spatial relations of joint co-existence in the same structure. We understand that the metaphors remain limited and urge the reader to beware of their unwanted spatial intimations. 

As for the objection from quantum non-locality to Humean supervenience, it can no longer even be stated at the fundamental level if the world is not fundamentally spatiotemporal. Of course, as long as spacetime is emergent we would expect there to also emerge non-local quantum connections. But non-locality makes no fundamental sense. Any problem arising from quantum non-locality now looks minor compared to the failure of Humean supervenience in non-spatiotemporal contexts. The urgent challenge for the Humean is now to find a way of characterizing the supervenience base in non-spatiotemporal terms while at the same time implementing the Humean credo that there are no necessary connections in nature in the new context. 

In fact, not only can Lewis's version of Humean supervenience no longer be maintained in the absence of spacetime, the worlds assumed to be unified and isolated by spatiotemporal relations fall apart as the spatiotemporal relations are no longer present to perform their duty of world making \citep{Wuthrich2020}. How can the Humean idea be salvaged into the spacetime-less context? To this end, we need to identify a fundamental natural external relation, which can function as world-making relation.

\subsection{Glue, but no nomic voltage}
\label{ssec:hpos}

We take two elements to be central to a Humean approach to laws (and, in fact, to much else). First, the supervenience base of occurrent facts. Second, a recipe for how to extract the laws from this base (or modality in some other form). 

As for the latter, the laws are presumed to supervene on, or be determined by the base, in the spirit of Physical Statism. This step in principle includes a full methodology of articulating the best system, and so to identify its axioms and theorems, i.e., the laws. Undoubtedly, there is much to say about it. Whatever its merits or problems, however, it is arguably not touched by the move to QG, where we assume scientific methodology to remain the same as in physics more generally. Consequently, it stays outside the scope of the present study.

The first step, however, changes dramatically if spacetime is not available at the fundamental level at which these facts are to be articulated. The supervenience base cannot consist in local states of affairs spatiotemporally arranged into a `mosaic', or even just a `carpet'. And it must satisfy two requirements. First, in order for the world to be unified (and perhaps isolated from other worlds), the particular contingent facts, or basic entities or states of affairs also require some form of `weave' or `glue' in order to form a world.\footnote{We wish to remain neutral as to whether the fundamental ontology consists in facts, substances, entities, or states of affairs. We believe that our argument does not depend on that choice.} Although we have just expressed the requirement in the Lewisian idiom, neither modal realism nor most other parts of the Lewisian system are presupposed: it is just that the facts in the base must cohere to form a whole. What kind of relations between the basic entities could take over this weaving or glueing job? Arguably, these relations should be appropriately external, i.e., they ought to be relations which the particular facts could stand in or not as a matter of contingent circumstance. In short, the relations should furnish the glue with which to connect the basis into a cohesive structure.

The second requirement is, of course, that there be no necessary connections between the independent facts or states of affairs in the basis, i.e., that the basis must be `categorical' in that proper parts of it do not necessitate anything about other parts.\footnote{By `categorical', we mean {\em unconditional}, in the sense that there are no dispositional or similarly modally latent aspects involved.} For the approach to remain appropriately Humean, then, the glueing connections cannot be modally or nomically `charged'. In short, the relations cannot introduce a form of `nomic voltage'. 

Although our target in this section so far might have appeared narrowly Lewisian, rather than more broadly Humean, we believe that at least these two conditions must be met by any Humean analysis to laws: we must start out from a cohesive basis free of necessary connections. 

In choosing such a basis, we may proceed more ontically, as does Lewis, by starting out from what there is, as some fundamental set of occurrent facts and see the laws supervene, perhaps because we believe in lean metaphysics. This is Lewis's vantage point, which is deeply rooted in metaphysics and in considerations regarding how one can construct a metaphysical system from the fundamental bottom up such that it maintains a Humean spirit in eschewing necessary connections. 

In contrast, and arguably closer to Hume's original perspective, we approach the problem in a more epistemic manner, by starting out from some set of facts which are accessible to us through experience---a manifold of evidence---in a way the laws are not, perhaps because we remind ourselves of the empiricist origins of Humeanism. Instead of the construction of metaphysically possible worlds, we start out from experience as our vantage point for theorizing. Despite these difference in perspective, both approaches may be rightly considered to be Humean as we use the term.

If we go down the epistemic route, there may be no immediate problem with the absence of spacetime from the fundamental ontology of the world. Our vantage point would be the world as it manifests itself to us through experience. Our experience and our evidence arguably are ordered spatiotemporally, and the totality of this evidence may function as our basis. Perhaps this route presents the most promising Humean strategy. In order to fully succeed, the Humean pursuing it must show how we end up with more than a naïve empiricist regularity theory if we do not start out from an ontic basis. Either way, we will not pursue this epistemic route here, also because quantum gravity is notoriously remote from our direct epistemic access to the world. Instead, let us consider the ontic strategy. 

On a naturalist approach to laws of nature, it seems appropriate that the basis, and with it the glueing relations, should be determined a posteriori. How to positively characterize the supervenience base will, in general, thus depend on the direction of future fundamental physics. Any attempt of doing so risks replacement by superior future theories. This fallibility of our metaphysical tenets is, however, an epistemic risk that a naturalist must accept. Let us briefly consider two examples of candidate quantum theories of gravity in order to study, by way of example, how such a supervenience basis may look like. The two examples are causal set theory\footnote{\citet{Dowker2013}; for an introduction aimed at philosophers and an argument that causal sets are significantly non-spatiotemporal, see \citet{wut12}.} and loop quantum gravity\footnote{\citet{RovVid15}; for an introduction aimed at philosophers and an argument that spin networks are significantly non-spatiotemporal, see \citet{Wuthrich2017}.}. 

Causal set theory postulates that what there is, fundamentally, a causal set. A causal set is a discrete structure consisting in otherwise featureless basal events, which are partially ordered by a relation of causal precedence. It is thus the causal relations which furnish the `glue'. This specifies what is often considered the `kinematic' structure of the theory. However, as this characterization is far too weak in that almost none of the ways a structure can satisfy the kinematic demands gives rise to worlds remotely like ours. Consequently, additional constraints must be imposed on the fundamental structure. Many of them can be considered `dynamic', in a potential analogy to more familiar physical theories. They are intended to describe a `birthing process' of how new elements can be added to an existing causal set, which is supposed to more generically lead to worlds at least with a cosmological structure like ours. On a more traditional view of laws, it is these `dynamical' principles---their detailed formulation need not detain us here---which ought to be considered the laws according to causal set theory. As the kinematical principles are also required as axioms of the theory, the kinematic structure of causal sets would also be nomic on a best systems analysis. 

The kinematic structures of loop quantum gravity are quantum superpositions of spin networks. A spin network is a combinatorial structure which can be represented by an abstract graph consisting in labelled edges connected by vertices, which also carry labels. Those edges connected by a vertex are interpreted to be `adjacent'. These adjacency relations thus deliver the `glue' connecting the edges to a cohesive structure. A spin network is in some sense only what is expected to give rise to spatial structure. In order to recover full space\textit{time}, some `dynamics' must be added. This step turns out to be technically non-trivial. There exist two different manners of completing the theory with a dynamics: either one subjects the spin networks to the so-called Hamiltonian constraint equation (the `canonical' way), or else one provides transition amplitudes between `initial' and `final' spin networks (the `covariant' way). Traditionally, one would deem either the Hamiltonian or these transition amplitudes to express the dynamics and so the laws of the theory. On a best systems analysis, the kinematic buildup concerning the spin networks will certainly have to be part of the axiomatic structure of the theory and also be considered nomic. 

But now we have a problem: the glue comes with nomic---and so with necessary---connections. For Lewis and other Humeans, the principle of recombination guarantees the absence of necessary connections between distinct entities: according to this principle, elementary entities (whatever they are) are \emph{freely} recombinable in the sense that any combination or permutation generates a metaphysically possible world. It is really this principle which underwrites the non-nomicity of the supervenience basis. 

Traditional philosophers identify the supervenience basis to be freely recombined pre-the\-o\-ret\-i\-cally. Not so for a naturalist who looks to quantum gravity: for them, theories guide us concerning what the basic facts are and the ways in which they can be recombined. Hence, any such naturalistically identified basis is nomically charged: the facts are \textit{not} freely recombinable, but only in accordance with the theory. 

In the two examples above, already the kinematical structures---the causal sets or the spin networks---are subject to axiomatic demands by the theory and so are steeped in laws. The connections they sanction are anything but free of necessity. In other words, to invoke a theory to license permissible recombinations amounts precisely to imposing laws on the basic facts, thus creating necessary connections.

It appears as if the naturalism looking to possibilities supported by fundamental physics ends up with a `nomically charged' supervenience base and so stands in tension with the Humeanism demanding that there be no necessary connections among particular matters of fact. Although the second step in the Humean project, the recipe to extract modality from a non-modal basis, may survive the loss of spacetime unscathed, the first step faces an existential threat from the naturalistic move towards cutting-edge physics: not only may spacetime vanish and with it the pre-theoretically fixed basis (at least \`a la Lewis), but turning to fundamental physics to identify a new supervenience base appears to make it impossible to end up with a base which is modally sufficiently innocent.

Consequently, it seems as if a Humean who seeks guidance in fundamental physics in identifying the base must reject the theory that guided them and insist that the identified structure can be freely recombined. Conversely, a naturalist who sticks to the guidance offered by fundamental physics must accept some necessary connections and so cannot maintain their Humeanism. A naturalist Humean, it seems, finds themselves in the awkward position of accepting what our best fundamental theories tell us what binds particular matters of facts (in order to obtain the `glue'), without accepting the theory `fully' (trying to avoid the attendant nomic voltage). 

A final point: one might think that the tension we identified here for a naturalist Humean arises before we get to quantum gravity and so is independent of it. Presumably, when we characterize a spacetime with general relativity or even just Newtonian gravity in mind, it seems as if we impose similarly nomological constraints on the supervenience base. Take the example of general relativity. Taking spacetime as fundamental posits may then be unproblematic, but we have to be careful not to exact conditions too strong from the spatial and temporal structure with which we constitute the supervenience base. For if it is overly restrictive, we run the risk that general relativity---which is in some ways a very permissive theory---accepts situations as possible that our overly strong conditions excluded. For instance, we might have thought that absolute simultaneity or the absence of causal loops are conditions that the spatiotemporal structure of our base must naturally meet, only to find out that general relativity admits physical possibilities in which these conditions are violated. In this case too, a naturalist might be tempted to require of that spatiotemporal structure that it accords with general relativity. But such a requirement would violate a free recombination and so imbue our base with necessity. The solution would of course be to impose only very weak a priori conditions on the spatiotemporal structure of the supervenience base. However, a naturalist would have to remain perennially open to the possibility that a superior future spacetime theory postulates hitherto forbidden spatiotemporal structures. And our conditions cannot be arbitrarily weak lest the remaining structures have nothing spatiotemporal about them. 

We accept that there is something like this tension present in naturalist Humeanism already before we get to quantum gravity. Previously, the tension was downplayed by the hope that spatiotemporal relations can be pre-theoretically conceptualized in appropriately weak terms, so as to avoid reference to a physical theory. However, quantum gravity throws it into the sharpest possible relief: if the fundamental structures quantum gravity postulates are not spatiotemporal anymore, then no prior weakening of the conditions could have hoped to antecedently give an acceptable characterization of the supervenience base.

\subsection{Navigating the tension}
\label{ssec:tension}

Is there any hope that this tension can be navigated? Although we believe that a conception of laws which identifies the laws of a theory with its \textit{dynamical} laws is unduly narrow, there is no doubt that dynamical laws are paradigmatic cases of laws in physics. This suggests a natural division between a scientifically informed framework for characterizing the supervenience base and the laws, which need not be imposed: the division between the kinematical and the dynamical structure of a physical theory. Thus, one could take the kinematically possible models of a theory as metaphysical possibilities and its dynamically possible models as nomological possibilities. In this way, the kinematics would furnish the glue, but the nomic voltage of the dynamics would be avoided in these merely kinematically possible worlds. 

This solution would surely be pleasing to the naturalist, as it departs directly from the basic structure of many physical theories in order to obtain a template for the metaphysical work. However, this method comes with some severe difficulties. First, the proposal presupposes a natural distinction between kinematics and dynamics of a theory. While there is such a distinction in remarkably many physical theories, this is not necessarily the case, and it can certainly not be presupposed in quantum gravity. For our two examples above, there is a sense in which the distinction is maintained. In causal set theory, one can regard the axiom that the basic structure is a causal set as articulating the kinematic framework and the additional conditions which are imposed as dynamical laws. Thus, any causal set is metaphysically possible and can serve as supervenience base, but only some of them are nomologically possible. In loop quantum gravity, the spin networks can be considered the kinematical possibilities, and the dynamical content as being captured either by the Hamiltonian constraint equation or the transition amplitudes. 

These examples clearly exhibit the second difficulty: the kinematic axioms specifying the structures of causal sets and spin networks are clearly also axioms of the theory and so laws in the perspective of the best systems analysis. For an advocate of this analysis, dynamical laws are not the only laws. Consequently, the proposal seems to be insufficiently Humean in that the kinematical possibilities used to furnish the glue are already nomically charged in that they impose nomic necessity on the supervenience base. 

Finally, there is a sense in which the proposal might be considered too narrowly naturalistic: whatever is not a kinematic possibility according to the chosen fundamental theory will not count as a metaphysical possibility. Thus, the proposal stays very close to the framework of a particular theory and so may not span the full gamut of what would generally be considered metaphysical possibilities. This may not be deemed a serious deficiency---and it is certainly a feature shared with Alastair \citet{Wilson2020}'s recent proposal of a naturalistic account of modality---, but it does limit the scope of the approach.

A different proposal maintains the suggestion that we start out from a space, but replace the traditional physical space with a different kind of space. In lieu of physical space, we look to the `state' space of configuration space of fundamental theories.\footnote{This move is similar to Loewer's proposal in the context of quantum mechanics (see \S \ref{ssec:hneg}). We thank Jonas Waechter for the suggestion.} Since the theories of interest will be quantum theories, we have a configuration space or a Hilbert space which can be postulated as fundamental. Thus, we use another space as a foundation for the Humean supervenience base, and the proposal inherits many of the Humean advantages of Lewis's account based on spacetime: it delivers a principled way to characterize the base and to identify possibilities. This approach is clearly naturalist as we start out from a physical theory, which of course has to earn its spurs in the usual scientific manner. Returning to our two examples, causal set theory is not yet sufficiently developed to the quantum level for this approach to gain much traction. In loop quantum gravity, the spin network form a basis in the  `kinematic Hilbert space'. Thus, for loop quantum gravity, this second approach would come to much the same as the first one. In fact, this is not unexpected given the general structure expected of a quantum theory.
 
While the proposal is promising, it also comes with its share of problems. First, it does seem to presuppose a space which may be too specific. In fact, every system has it own state or Hilbert space. As a consequence, we find that for example the dimensionality of these spaces depends on the number of degrees of freedom. Unlike spacetime, if we add a single particle, these spaces change. Thus, fixing a Hilbert space, say, would be very restrictive in that we would obtain, at best, a free recombination for that particular system only. Of course, that system can be the entire universe. But even in that case, there is a sense in which this approach would be committed to accepting that there could not have been a particle more or less in the universe. A possible response to this challenge could be not to consider a single, fixed Hilbert space, but instead a space consisting in a whole range of Hilbert spaces---one for each total system that the theory deems possible. 

Second, even if there is a way to generalize the approach to resolve the first difficulty, it would still make free recombination semi-hostage to a particular theory (or perhaps family of theories), leading back to the tension between naturalism and Humeanism. 

Finally, an important proposal offered by \citeauthor{Jaksland2021} (forthcoming) is to take entanglement as the world-making relation for an ontology of `systems': two systems are worldmates if and only if they are quantum-mechanically entangled. Following Jaksland's proposal, entanglement could thus replace distance or cognate notions as the glue we are seeking. Given that we would generally expect that a world can be partitioned into (sub)systems, and that these would generically be entangled with the rest of the world, entanglement---of course purified from any spatial connotations---appears as a promising candidate relation to make worlds. And since an appropriate theory of quantum gravity would be a \textit{quantum} theory, Jaksland's proposal remains an appealing option also in quantum gravity. What needs to be established for that option to be viable in quantum gravity is that the fundamental entities of quantum gravity, whatever they turn out to be, somehow amount to an ontology of `systems' which are generically entangled with one another. 

This is arguably the most promising proposal. However, it should be clear that here too we are banking on some particular feature expected in a future theory of quantum gravity to deliver the fundamental glue, thereby ruling out all other kinds of worlds devoid of this particular way of glueing parts into worlds.\footnote{More specifically, the proposal also seems to rule out simple worlds in which there are no parts which can be entangled with one another. Perhaps this is an acceptable cost.} And similarly to the other proposals above, the Humean needs to argue that the theoretical constraints (e.g.\ on recombinations) linked to the entanglement glue holding the basis together are not nomically charged in any worrying sense.

We have now briefly discussed three proposals to identify the fundamental glue of worlds: merely kinematic structures, structural relations in Hilbert spaces or similar, and entanglement. All these proposals come with serious naturalistic credentials and deserve to be pursued in further research. We will not here advance any concrete proposal along the ontic route to a full account of the supervenience base, hoping that the discussion above illustrates the fundamental tension with naturalism any such Humean approach faces. Perhaps the difficulties we are facing here speak in favour of returning to the epistemic approach of identifying the supervenience base and build out Humeanism from a manifold of experience. But this is the project for another day.

Or perhaps the way to see the tension eased is by rephrasing the tension as a dilemma. Either we are naturalistically inclined recombinatorialists, in which case we just have to accept to pursue the best option from the list above and accept its implications. Or else we are modal realists of the Lewisian bent, in which case recombination is sufficient to generate a range of worlds, but it is not necessary in the sense that the generated worlds are not expected to be exhaustive of all metaphysical possibilities. In other words, we adopt a form of contingentism and happily embrace possible worlds beyond those sanctioned by fundamental physics. Either way, the tension is solved. In response, we should note that we accept a severe restriction to the range of possibilities our analysis can account for and so restrict our Humeanism on the first leg, or we give up much of our naturalism on the second leg. This is precisely what the tension consists in.

\section{Primitive modality without spacetime}
\label{sec:pd}

We now turn to the two main non-Humean conceptions of (fundamental physical) laws, namely primitivism and dispositionalism. As its name suggests, primitivism holds that (fundamental physical) laws of nature are ontological primitives, part and parcel of the fundamental ontology of the world (\citealt[ch. 2]{Maudlin2007} is a paradigmatic example). According to dispositionalism, the laws are grounded in the fundamentally dispositional or causal nature of properties, i.e., in fundamental dispositions (see \citealt{Bird2007}) or causal powers (see \citealt{Shoemaker1980}). These characterizations are rather rough and there are many metaphysical differences (and subtleties) among the various different versions of primitivism and dispositionalism, but these are not directly relevant for our focus on the spacetime aspects of these conceptions. In this perspective, the crucial common feature of primitivism and dispositionalism is that they involve some form of (irreducible) primitive modality giving rise to the spatiotemporal distribution of particular facts---both conceptions are non-reductive with respect to some modality (or with respect to the laws themselves), in stark contrast to the reductive Humean view. Within this framework, the entire Humean mosaic need not be accepted as primitive---postulating the mosaic from the start is very unsatisfying from the non-Humean point of view, since it is then not accounted for. Instead, it is derived from a proper subset of particular facts endowed with some primitive modality, which is either conferred by the primitive laws or grounded in fundamental dispositional properties. 

In a first step, this section aims to identify the difficulties that primitivism and dispositionalism about laws face in the context of quantum gravity where space and time may not be (fully) present. Prima facie, it seems straightforward to adapt these conceptions to a non-spatiotemporal setting: the novel laws of quantum gravity just need to be taken as primitive or as being grounded in the (stipulated) fundamental dispositional nature of the relevant quantum gravity degrees of freedom. However, beyond mere stipulation, these conceptions are usually motivated and articulated against a temporal background that may not be available in quantum gravity---indeed, they are actually in tension with the dynamical nature of spacetime already at the classical level of general relativity (\S\ref{ssec:pdneg}). In a second step, we will attempt to specify some generalized, non-temporal notion of `nomic production' which lies at the heart of both primitivism and dispositionalism in the context of the concrete examples we have considered in the last section, namely causal set theory and loop quantum gravity (\S\ref{ssec:pdpos}). 

\subsection{The temporal background of primitivism and dispositionalism}
\label{ssec:pdneg}

Both primitivism and dispositionalism fundamentally encode temporal features in their standard formulation. Primitivism revolves around the conception of laws governing the temporal evolution of some physical state. Similarly, within dispositionalism, laws are anchored in the very nature of properties that are understood in terms of irreducible powers or dispositions to produce certain effects.

When it comes to fundamental physical laws, these temporal aspects are made explicit through the natural articulation of primitivism and dispositionalism within the framework of the initial value formulation of physical theories. In this context, the ontological picture is that the spacetime distribution of particular facts is not primitive, but produced and determined by the initial conditions of the world together with the primitive fundamental laws or by the fundamental dispositional properties instantiated by the initial state of the world. This picture is made very explicit in \citet[ch. 6]{Maudlin2007}, where he writes that, according to his primitivism, ``the total state of the universe is, in a certain sense, derivative: it is the product of the operation of the laws on the initial state'' (182).\footnote{Throughout this section, Maudlin's primitivism (as exposed in \citealt{Maudlin2007}, and more recently in \citealt{Maudlin2020}) will often serve as an illustration, partly because it explicitly captures features that are common to other non-Humean conceptions of laws, and also because it has been articulated with fundamental physics in mind.} 

The dispositionalist conception about laws and properties is very similar in this respect: in virtue of their irreducible dispositional essence or causal powers it is in the very nature of the fundamental physical properties to produce certain effects, so that these latter can be considered as derivative. In this sense, for both the primitivist and dispositionalist, the entire spacetime distribution of particular facts (`the total state of the universe')is the necessary consequence of the initial conditions of the world (plus the laws in the primitivist case). 

The notion of `production' here---in the case of primitivism, also: `governance'---is therefore central to the non-Humean nature of both primitivism and dispositionalism. The above standard characterization of these conceptions has obvious temporal aspects, since `the total state of the universe' is grounded in its \emph{initial} state (prima facie, `production' actually just is a temporal notion)\footnote{See \citet[175]{Maudlin2007}: ``The basic temporal asymmetry of past-to-future underlines the very notion of production itself, so that without it there can be no production.''\label{ftn-tempasym}}. In many ways, time is a fundamental dimension in which the laws operate (`govern') or in which the dispositions or powers manifest themselves, producing their effects. Indeed, \citet[ch. 6]{Maudlin2007} explicitly takes time and the passage of time as ontologically primitives, part of his ``Non-Humean Package'' (182) along with the fundamental laws.\footnote{The situation is a bit less explicit in the literature on dispositionalism, partly because it is in general more metaphysical and tends to pay less attention to physics than does Maudlin's primitivism.} 

This primitive temporal background, against which the laws operate, plays a crucial role within the primitivist framework, in particular when it comes to physical possibility, counterfactuals and explanatory power. The case of counterfactuals within the framework of Maudlin's primitivism is especially telling, since it heavily relies on the initial value formulation of fundamental laws. As an example, \citet[22-23]{Maudlin2007} considers the following counterfactual statement: ``If the bomb dropped on Hiroshima had contained titanium instead of uranium it would not have exploded.'' The initial value formulation of the relevant laws and the corresponding temporal structure allows Maudlin to evaluate and to ground the truth value of this counterfactual through a three-step process (\citeyear[22-23]{Maudlin2007}): ``Step 1: choose a Cauchy surface that cuts through the actual world and that intersects the bomb about the time it was released from the plane. All physical magnitudes take some value on this surface. Step 2: construct a Cauchy surface just like the one in Step 1 save that the physical magnitudes are changed in this way: uranium is replaced with titanium in the bomb. Step 3: allow the laws to operate on this Cauchy surface with the new boundary values generating a new model. In that model, the bomb does not explode. Ergo (if we have got the laws right, etc.) the counterfactual is true.'' The truth value of the counterfactual is grounded in the \emph{temporal} (Cauchy) development (governed or produced by the laws) of the suitably modified Cauchy data (according to the antecedent of the counterfactual). So, the temporal background---the temporal structure of the initial value formulation---is clearly playing an important role here in the work done by the laws (the importance of this role is nicely illustrated within the framework of Maudlin's primitivism, but we take our considerations above to suggest that the situation is similar in this respect in the case of dispositionalism, at least when it comes to fundamental dispositional properties). 

At this stage, it is interesting to note that the initial value formulation has also recently been exploited by the Humean camp. Indeed, \citet{Hall2015} effectively uses the tools of the initial value formulation to sharpen the best system analysis (see \S \ref{ssec:hneg} above).\footnote{It should be noted though that \citet{Hicks2018} has contested that the orthodox best systems analysis has the resources to distinguish between dynamical laws and initial conditions.} The framework of the initial value formulation (with its temporal structure) allows him to specify more precisely the notoriously controversial Lewisian criteria of simplicity and strength in terms of what he calls an ``initial condition hypothesis'' and a ``dynamical hypothesis''. The exact details are not relevant here, but what is interesting is that \citeauthor{Hall2015} explicitly introduces the temporal (and spatial) structure of the initial value formulation into the best system analysis, because he acknowledges its crucial role in our understanding of fundamental laws and of what they do (in particular about counterfactuals).

Now, it is important to highlight that the temporal structure underlying the initial value formulation and (more or less explicitly) assumed in the primitivist and dispositionalist conceptions of laws is already in tension with certain central features of general relativity. More precisely, the background spacetime structure required by the initial value formulation is at odds with the dynamical nature of spacetime encoded in general relativity.\footnote{The dynamical nature of spacetime within general relativity creates difficulties for Maudlin's procedure for evaluating counterfactuals, which can be precisely specified in terms of non-linearity and ellipticity within the framework of the initial value formulation (\citealt{Jaramillo-Lam2021}).} Indeed, one of the crucial implications of the dynamical nature of spacetime is that its global topology is not fixed a priori, and so may not allow for a well-defined initial value formulation (existence and uniqueness theorems for the initial value formulation of general relativity assume global hyperbolicity, which amounts to an imposition of the global $3+1$ spacetime topology $\mathbb{R} \times \Sigma$, where $\Sigma$ is any Cauchy surface). As a consequence, primitivism and dispositionalism about laws as articulated above can be specified only within a certain class of general relativistic spacetimes---mainly: globally hyperbolic spacetimes.\footnote{The initial value formulation can actually be extended to certain non-globally hyperbolic spacetimes (\citealt{Friedman2004}).}  

Similarly, general relativistic spacetimes may allow for closed timelike curves (as for instance within the Kerr-Newman solution, describing spacetime around a rotating charged mass), which create difficulties for the very notion of `production' at the heart of the non-Humean nature of the primitivist and dispositionalist conceptions of laws. \citet[174-175]{Maudlin2007} is very explicit about the fact that closed timelike curves would undermine the sort of productive explanation of the Humean mosaic that (his version of) primitivism about laws provides: ``Production is clearly transitive, so production-in-a-circle would imply self-production, but an item cannot be the \emph{ontological} ground of its own production. This means that the existence of closed timelike curves would imply the non-existence of this sort of productive explanation, and might suggest that a Humean account is the strongest that can be had.'' What is interesting for us here is that this stance further underlines central features of the non-Humean strategy at work in the primitivist and dispostionalist conceptions: that is, to consider a (proper) subset of all the occurrent facts (i.e. a subset of the Humean mosaic) as the ontological ground for the rest of the totality of occurrent facts (i.e. for the whole Humean mosaic), the latter then being \emph{produced} and therefore \emph{productively explained} by the former. This strategy is explicitly non-Humean in that it encodes some primitive modality---in the primitive laws or in the irreducible dispositions---as well as some form of necessity---between the ontological ground and what is derived from it. As we will see in the next section (\S \ref{ssec:pdpos}), one way for the primitivist and dispositionalist conceptions of laws to face the challenge that arises in the context of quantum gravity is to articulate and generalize this non-Humean strategy in a way that does not rely on spacetime and on spatiotemporal notions.     

Of course, within the framework of general relativity, the primitivists and the dispositionalists can simply restrict the scope of their account to globally hyperbolic spacetimes, which do not contain closed timelike curves and where the initial value formulation is well-defined; for instance, within Maudlin's primitivism, closed timelike curves are considered to be (meta)physically impossible. Despite the fact that it can be justified (to some extent) by cosmological considerations about our actual universe being globally hyperbolic (at least in a certain approximation), such a strong metaphysical attitude is not very satisfying from a naturalistic point of view, since it amounts to constraining a physical theory with metaphysical prejudices (about the need for some fixed spatiotemporal structure) while refusing to inform the metaphysical framework with physical insights (about the dynamical nature of spacetime).\footnote{To be fair, \citet[48]{Maudlin2007} does consider the possible need of revising his conception of laws in the light of new physics (``[f]uture theories may challenge the notion of time itself'').} In any case, this attitude is not tenable in the context of quantum gravity where spacetime is not fundamental.

\subsection{Non-temporal nomic production}
\label{ssec:pdpos}

It is obvious that the standard expression of the primitivist and dispositionalist conceptions of laws relying on the initial value formulation cannot even be properly articulated if the relevant spacetime structures are absent. As we have seen in the last section (\S \ref{ssec:pdneg}), these difficulties for primitivism and dispositionalism are not merely the consequences of wild theoretical speculations put forward by various research programs in quantum gravity, but actually find their roots already at the classical level in central features of the well-established (and experimentally successful) theory of general relativity---specifically: in the dynamical nature of the spacetime structure itself, a feature that lies at the heart of major research programs in quantum gravity, such as loop quantum gravity and causal set theory.

In the case of primitivism about laws, the difficulties can apparently easily be overcome at first sight, since quantum gravity laws or nomic relations, whatever they are according to the theory, just need to be accepted as primitive---after all, as \citet[18]{Maudlin2007} puts it, the primitivist ``analysis of laws is no analysis at all''. However, on its own, such a trivial move is hardly satisfactory, even by the primitivist's own lights, since primitivism about (fundamental physical) laws (at least in Maudlin's version) is justified by the work (fundamental physical) laws can do, if accepted as primitive, in providing a clear foundation for notoriously difficult notions such as counterfactuals and causation. But it is not clear what exactly it is that accepting laws as primitive can achieve in a context without spacetime. For instance, the powerful primitivist scheme for evaluating counterfactuals we have discussed in the last section (\S \ref{ssec:pdneg}) heavily relies on the initial value formulation and so is just not available at the level of quantum gravity---at least, not in any obvious, straightforward way. 

We suggest that a more fruitful way to articulate primitivism and dispositionalism about laws in the context of quantum gravity is to specify a proper subset of all the fundamental quantum gravity facts or degrees of freedom as the ontological ground for the totality of quantum gravity facts, in virtue of primitive laws or irreducible dispositions. The idea is that certain fundamental facts endowed with some irreducible dispositions or together with some primitive laws would give rise---produce---in some \emph{non-temporal} sense the totality of quantum gravitational facts. Such a conception would encode both the fundamentally non-Humean intuition of some primitive modality as well as the fundamentally non-Humean motivation of some (productive) explanation for the non-spatiotemporal `distribution' of fundamental facts or degrees of freedom (hence avoiding the brute Humean acceptance of this entire distribution as being primitive, as within the Humean mosaic picture). In this context, primitive modality plays some sort of `glueing' role, fixing what is physically (nomically) possible, and, in a sense, holding things together. Thus, one of the main challenges of the Humean approach to laws without spacetime just does not arise here, since some primitive nomic glue is just assumed from the start. However, there is no free lunch: in the context of quantum gravity, primitivism and dispositionalism about laws face the difficulty (unknown to the Humeans) to articulate a non-temporal notion of nomic production---or in terms of \S  \ref{ssec:hpos}: the `nomic voltage' need to be specified in a non-spatiotemporal way.

In the naturalistic perspective adopted in this paper, the specification of such non-temporal nomic production cannot be an a priori matter, but need to be articulated within the framework of concrete approaches to quantum gravity. We consider the two examples of section \ref{sec:hume}: causal set theory and loop quantum gravity. In this context, one natural move for primitivism and dispositionalism is to exploit the distinction between the kinematical and dynamical parts of these theories: the latter being ontologically grounded or nomically produced (in some non-temporal way) by the former, which is either endowed with irreducible dispositions or accompanied by some primitive laws. 

For instance, within causal set theory, the full causal set underlying the entire spacetime (ultimately, the entire universe including its material or non-gravitational content) would be grounded in some primitive causal set---a proper subset of the full causal set---, in the sense that the former would be nomically produced from the latter through the fundamental causal set dynamics, as captured by, e.g., the (ultimately, quantum) sequential growth dynamics. In the primitivist and dispositionalist perspective, this dynamical law would simply be taken as primitive or as the manifestation of the fundamental (probabilisitic) disposition instantiated by the primitive causal set. It is important to stress here that the primitive causal set is not `initial' in any standard temporal sense and the dynamics itself is not explicitly temporal, since causal set theory lacks the relevant spacetime structures. In this sense, the above characterization provides some hints about how the primitivist and dispositionalist conceptions of laws can be generalized to a context where central spacetime structures are absent.  

Such a generalization can be further exemplified within the framework of loop quantum gravity, where the primitive kinematic combinatorial (spin network) structures can be similarly endowed with fundamental (probabilistic) dispositions or primitive dynamical laws that produce the full combinatorial (spin foam) structures underlying the entire spacetime (ultimately, the entire universe including its material or non-gravitational content). In this (covariant) context, the dynamics is typically captured by appropriate transition amplitudes associated with boundary spin network states, and is not temporal in any standard sense.      

The non-Humean strategy at play is very similar in these two examples in that it heavily relies on a non-spatiotemporal distinction between kinematical and dynamical structures: indeed, this distinction seems to allow the primitivists and dispositionalists to articulate a primitive, non-temporal notion of nomic production, thus encoding some primitive nomic necessity at the level of quantum gravity. So, from these preliminary considerations, it might seem that primitivism and dispositionalism about laws are more easily amenable than Humreanism to a context without spacetime. Great care is required though: indeed, a slightly more detailed analysis reveals several difficulties. 

A first difficulty arises as the very notion of `production' seems to involve some form of asymmetry in one way or another; as we have seen in \S \ref{ssec:pdneg}, that's precisely one of the principal reasons for \citeauthor{Maudlin2007} to consider the direction of time as fundamental (see footnote \ref{ftn-tempasym}). But such primitive temporal directionality is obviously not appropriate in quantum gravity and there is no guarantee that a non-temporal asymmetry obtains in this context. Within the framework of causal set theory, sequential growth---the `birthing process' of new elements---may well encode some not explicitly temporal asymmetry (which may help to ground some generalized version of becoming, at least within classical sequential growth---not without its own set of challenges though, see \citealt{Earman2008} and \citealt{Wuthrich-Callender2017}).

The situation is far less clear in the context of loop quantum gravity, where the theoretical apparatus does not explicitly display any such asymmetry. The transition amplitudes capturing the dynamics can be understood in terms of processes between `initial' and `final' spin network states, but `initial' and `final' are only labels in this context, without any standard temporal meaning (in the covariant perspective, one rather considers boundary spin network states). However, this does not mean that the situation is entirely hopeless for the non-Humean notion of production in loop quantum gravity, since the primitivist or the dispositionalist may still have several options available: for instance, they may articulate spin foam transition amplitudes in terms of non-temporal productive processes or look for additional quantum dynamical features involving some relevant fundamental asymmetry, such as within spontaneous dynamical collapse theories. So, much work remains to be done for the primitivist and dispositionalist about laws in this context and much also depends on the future developments of the physics, but as things stand now, it seems fair to say that there is no straightforward way to articulate a productive conception of laws in loop quantum gravity.  

Another difficulty for the non-Humean strategy above is the very distinction it relies on between the kinematical and the dynamical parts of quantum theories of gravity. As we have discussed in \S \ref{ssec:tension}, even if central in many physical theories (including research programs such as causal set theory and loop quantum gravity), this distinction cannot be presupposed to obtain or to remain meaningful in the context of quantum gravity.\footnote{Indeed, to some extent, the very dynamical nature of spacetime (and the related background independence) can be argued to blur the distinction between kinematics and dynamics already at the level of general relativity, see 
\citet[91]{Ehlers2007}
.} More generally, it seems that the explanatory virtue of the primitivist and dispositionalist strategy discussed here---which has the ambition of explaining how the Humean mosaic is produced from some ontological bedrock, instead of merely accepting it as primitive---relies on the existence of structures mimicking certain distinctively temporal features, such as some temporal asymmetry or directionality.\footnote{Primitivism and dispositionalism about laws may be hard to articulate in a fully spatiotemporal context where such asymmetry would be absent, as discussed in \S \ref{ssec:pdneg} in the context of general relativity.} These primitivists and dispositionalists may feel partly vindicated by the fact that, to some extent, some very basic distinction between `space' and `time' remains present in most (if not all) approaches to quantum gravity to date (\citealt{LeBihan-Linnemann2019}): for instance, loop quantum gravity explicitly implements local Lorentz invariance (which encodes some distinction between `space' and `time', as manifested in the Lorentz signature), whereas in causal set theory some distinction is encoded in the partial ordering structure (one may of course debate between what exactly the distinction obtains). However, tying the non-Humean conceptions of laws to such a distinction make them hostage to the obtaining of this distinction at the level of quantum gravity, something which of course cannot be guaranteed a priori. It should be noted that, as such, being vulnerable to future developments of physics is in any case a direct consequence of the naturalist attitude adopted here.

\section{Conclusion}
\label{sec:conc}

We have articulated the difficulties that the standard Humean, primitivist and dispositionalist accounts of laws face in the context of quantum gravity where central spacetime features---if not spacetime itself---are not fundamental, but only emergent. To some extent, the existence of such difficulties is to be expected, since they find part of their roots already at the level of quantum mechanics and general relativity, the ingredient theories of most approaches to quantum gravity. Indeed, these difficulties can be partly traced back to the tension between, on the one hand, central physical features in these theories (Bell-type correlations in quantum mechanics and the dynamical nature of spacetime in general relativity) and, on the other hand, the spacetime characterization underlying the standard analyses of laws. If the tension can be managed to a certain extent at the level of the ingredient theories, we have seen that the challenge becomes unbearable in the context of quantum gravity. The situation for the standard accounts of laws is, however, not entirely hopeless: we have suggested various ways of reconceiving them without relying on spacetime structures, identifying the challenging elements along the route. 

A central difficulty for the Humean resides in the articulation of fundamental non-spatiotemporal glueing relations that hold the basis together without being modally or nomically charged (so that they can be accepted as brute facts about the basis). Quantum entanglement and more specifically Bell-type correlations may play such a role, at least to the extent that they are central features in many approaches to quantum gravity.\footnote{There are also several suggestions about the role of entanglement in the emergence of spacetime, in particular some of them relying on (or inspired by) the Ryu-Takayanagi formula (or conjecture) in the context of AdS/CFT correspondance between entanglement entropy (in the boundary) and area (in the bulk), see, e.g., \citet{Ryu-Takayanagi2006} and \citet{VanRaamsdonk2010}; in the philosophical literature, see \citeauthor{Jaksland2021} (forthcoming), and for a critical perspective see \citet{Ney2021}.} As an example, there is a precise sense in which entanglement plays a glueing role among fundamental degrees of freedom in loop quantum gravity (\citealt{Baytas-et-al2018}); in this example, the Humean then needs to argue that the attendant glueing conditions or constraints (e.g.\ on the quantum polyhedra corresponding to the spin network nodes to be glued together) are sufficiently categorical, that is, sufficiently modally and nomically innocuous to be accepted as brute facts about the basis. In any case, what these conditions and their implications really mean for the Humean needs to be carefully assessed; and similarly for other glueing proposals---this is work for future research.

To some extent, the situation is even more challenging for the primitivist and dispositionalist conceptions about laws. We have seen that in this case one of the central difficulties is to articulate a meaningful notion of non-temporal production. The challenge is made even harder by the fact that the dynamics is in general the less well-known (and the most problematic) part in most of the approaches in quantum gravity.

In the face of these difficulties, one option is to adopt a pragmatic attitude towards the standard accounts of laws. In this sense, one could consider that, to the extent that spacetime emerges at some level, the standard, spacetime-based laws also emerge.\footnote{The framework of functional emergence developed for spacetime in \citet{lamwut18} may be of some help here.} Consequently, standard analyses of laws would not be applicable at the fundamental level (where, in some sense, there would be no laws), but instead at the derivative levels of known and established physics. Moreover, if according to our best theories of quantum gravity it is physically possible  that the fundamental degrees of freedom combine such that spacetime does not emerge, this approach would be committed to accepting that in such cases, there are no laws of nature at all. Of course, either way, such an attitude would simply evade the main question we have started with, namely the nature of laws in the context of quantum gravity. Or perhaps this is precisely the lesson to take from quantum gravity: that there may be no such thing as laws of nature, fundamentally.

%
%

\bibliographystyle{apa}
\bibliography{Ref_Laws}

\end{document}